\documentclass[sn-apa]{sn-jnl}

\newcommand{\oneS}{\ensuremath{{}^{\textstyle *}}}
\newcommand{\twoS}{\ensuremath{{}^{\textstyle **}}}
\newcommand{\threeS}{\ensuremath{{}^{\textstyle **}\oneS}}

\jyear{2022}%
\usepackage{lineno}
\usepackage[export]{adjustbox}

\theoremstyle{thmstyleone}%
\theoremstyle{thmstyletwo}%
\theoremstyle{thmstylethree}%
\begin{document}

\title{
Which Factors are associated with Open Access Publishing? A Springer Nature Case Study}


\author*[1]{\fnm{Fakhri} \sur{Momeni}}\email{Fakhri.Momeni@gesis.org}

\author[1,3]{\fnm{Stefan} \sur{Dietze}}\email{Stefan.Dietze@gesis.org}

\author[1]{\fnm{Philipp} \sur{Mayr}}\email{Philipp.Mayr@gesis.org}

\author[2]{\fnm{Kristin} \sur{Biesenbender}}\email{k.biesenbender@zbw.eu}

\author[2]{\fnm{Isabella} \sur{Peters}}\email{i.peters@zbw.eu}

\affil*[1]{\orgdiv{KTS}, \orgname{GESIS -- Leibniz Institute for the Social Sciences}, \orgaddress{\street{Unter Sachsenhausen 6-8}, \city{Cologne}, \postcode{50667}, \country{Germany}}}

\affil*[2]{\orgdiv{Web Science}, \orgname{ZBW -- Leibniz Information Centre for Economics}, \orgaddress{\street{Düsternbrooker Weg 120}, \city{Kiel}, \postcode{24105}, \country{Germany}}}

\affil*[3]{\orgdiv{Computer Sciences}, \orgname{Heinrich-Heine-University Düsseldorf}, \orgaddress{\street{Universitätsstr}, \city{Düsseldorf}, \postcode{40225}, \country{Germany}}}

\abstract{Open Access (OA) facilitates access to articles. But, authors or funders often must pay the publishing costs preventing authors who do not receive financial support from participating in OA publishing and citation advantage for OA articles. OA may exacerbate existing inequalities in the publication system rather than overcome them. To investigate this, we studied 522,411 articles published by Springer Nature. Employing correlation and regression analyses, we describe the relationship between authors affiliated with countries from different income levels, their choice of publishing model, and the citation impact of their papers. 
A machine learning classification method helped us to explore the importance of different features in predicting the publishing model. The results show that authors eligible for APC waivers publish more in gold-OA journals than others. In contrast, authors eligible for an APC discount have the lowest ratio of OA publications, leading to the assumption that this discount insufficiently motivates authors to publish in gold-OA journals. We found a strong correlation between the journal rank and the publishing model in gold-OA journals, whereas the OA option is mostly avoided in hybrid journals. Also, results show that the countries' income level, seniority, and experience with OA publications are the most predictive factors for OA publishing in hybrid journals.}


\maketitle

\section{Introduction}\label{sec0}
The unrestricted availability of Open Access (OA) publications is linked to the goal of granting all interested parties free access to scientific knowledge and ensuring greater equality of access \citep{munafo2017manifesto}. This view is strongly related to the consumers of scholarly knowledge, who then would not have to pay for access. However, when taking the authors of those articles into account, they are affected by OA in two different ways: a) when choosing a publication model for an article and b) when receiving citations (and along with its reputation) for articles that have been published via a certain model (usually described as citation advantage\citep{langham2021open}). Those two aspects of OA may introduce significant biases and inequity into the scholarly publication and reputation system since they may restrict participation in OA in particular ways \citep{bahlai2019open}. 

First, the OA publishing model generally shifts the publishing costs from readers to authors or their institutions and funders by introducing article processing charges (APCs). This can be a severe constraint for those authors who cannot afford these costs or do not receive any financial support. To overcome this issue, most publishers have implemented an APC waiver/discount policy for authors from, e.g., low-income countries \citep{lawson2015fee}. However, it is open how the different options for OA publishing and waivers/discounts are considered and adopted by researchers with various characteristics such as their countries’ income level, but also their seniority and gender – factors which are also often associated with the decision to publish OA \citep{iyandemye2019low,olejniczak2020s,simardgeographic,smith2020assessing,zhu2017support}. \cite{rouhi2022left} discussed the waiver issues from the perspectives of the publisher, institutions, and developing countries. They mentioned the potential unfairness authors are confronted with, which may be caused by APC-based models. They argued that waiver programs have yet to address this problem successfully. They suggested that meeting the equity standard requires a cross-functional approach involving publishers, funders, research institutions, individual researchers, libraries, and service providers.  

To accommodate OA publishing costs, three funding options have emerged over time. First, Diamond OA journals are funded by public institutions such as libraries, which enable free reading and publishing for all researchers. Second, transformative agreements between public institutions and publishers have been introduced that include reading and publishing contracts and which are also funded by the institutions. In this case, there are no direct fees for authors, but their institutions pay for the APCs as part of a consortium. Access to publishing and access to publications is limited to participating organizations only. Thirdly, APCs could also be paid by the authors or their institutions themselves. The first option leads to Gold OA at the journal level. Transformative agreements allow authors to publish in either gold OA or hybrid (which -- for a fee -- allow publishing individual articles as an OA-variant) journals. The third option is often associated with hybrid journals. All other publishing models for journals usually require funding via subscriptions, resulting in closed-access articles (CA) that can only be read after paying the article or journal fee. 

The publishing model is also strongly associated with the visibility of authors and articles. For many researchers, it makes a difference where, i.e., in which journals they publish (e.g., considering discipline-specific journal rankings). If they want to be noticed by others and/or seek promotion, it can be crucial to publish in reputable journals, especially for early career researchers. And to achieve this, not only financial hurdles and APCs have to be overcome, but, for example, English language skills and technical skills are needed, as well as institutions that can help with legal advice or infrastructure support. Against this background, researchers have to decide which publishing model to choose and whether OA is not only an altruistic but feasible option at all.

The second possible source of bias and inequity is related to the paying for access case: It has been shown already that articles published as OA-variants are more visible, leading to higher citation counts and altmetrics \citep{evans2009open,lewis2018open,mckiernan2016point,ottaviani2016post,fraser2020relationship}. Moreover, the Matthew effect shows that researchers who are already well-known and widely cited receive even more citations \citep{farys2021matthew} -- which directly affects rewards for publication in prestigious journals, for prominence, and citations. For researchers, publications play a central role in their daily practice and the reputation system in which they operate. Publications enable researchers to build on the body of knowledge and refer to those findings by citing the publications (which accumulate reputation in this way). Hence, access to publications is crucial for the progress of science and building of reputation – which both can be impeded by a lack of access to OA publishing options and the risk of CA-articles not being cited as frequently as OA articles.

From that, we hypothesize that researchers with better access to financial resources have better access to publications – both in terms of access to read openly and in terms of access to publish openly. Associated with that may be an even stronger citation advantage for those researchers (usually WEIRD: Western, educated, industrialized, rich, and democratic; \citep{henrich2010weirdest}) with extensive OA-publishing options. As such, OA may carry the risk of perpetuating already existing inequalities rather than resolving such marginalization in the scholarly communication system \citep{fox2021open}.

 \section{Related work}
 Related work also indicates a strong association between economic factors, OA, and citation advantages. The scientific output of countries is associated with their economic evolution because scientific progress needs governments' financial support. \cite{samimi2011scientific} used a Granger Causality Test to examine the causal relationship between scientific output and GDP in 176 countries and found a two-way positive relationship between them. \cite{king2004scientific} compared published papers and their citation impacts across countries and found that only 31 countries contributed to 98\% of the world's highly cited papers and that the remaining 161 countries contributed less than 2\%. 
 
Open Access publishing is also highly influenced by the authors' country of affiliation since it determines APC waiver/discount policies or the availability of transformative agreements with publishers.   
Some publishers offer general waivers or have a discount policy for all of their journals for eligible authors, and the country's income level mainly determines eligibility.  \cite{lawson2015fee} has studied the waiver policy of the 32 most prominent publishers and found that 68\% of them grant APC waivers.  
\cite{simardgeographic} found that low-income countries publish and cite OA more than upper-middle and high-income countries. The positive correlation between OA citing and publishing is 1.3 times weaker for high-income countries than other countries. Similarly, \cite{iyandemye2019low} showed that biomedicine researchers from low-income countries have the highest percentage in OA publishing. \cite{smith2020assessing} reported the proportionately fewer OA articles published in Elsevier's journals for low-income countries, despite their eligibility for APC waivers.

\cite{olejniczak2020s} studied the articles published by faculty members at research universities in the United States and found that in the United States, male and senior authors are more likely to publish in OA form. \cite{zhu2017support} conducted a survey with over 1800 researchers at 12 Russell Group universities\footnote{\url{https://russellgroup.ac.uk/about/our-universities/}} to find the differences in OA publishing regarding discipline, seniority, and gender. Their results revealed disciplinary differences in OA publishing (Medical and Life Scientists are most likely to publish in Gold OA journals), more tendency toward OA publishing for senior authors, and across genders for men.

The journal rank is a decisive factor in submitting the article in addition to its business model. \cite{schroter2005perceptions} conducted a survey study with 28 international authors who submitted to the BMJ and found that for authors, the journal's ranking is more important than the availability of OA.

Many studies have investigated the OA citation outcome, and most found a citation advantage for OA articles \citep{evans2009open,fraser2020relationship,lewis2018open,mckiernan2016point,ottaviani2016post}. However, regarding biases (e.g., quality bias, self-selecting, mandating, self-archiving), different sampling and controlling data makes it difficult to conclude that receiving more citations is only the effect of OA. \cite{momeni2021happens} studied the citation impact of flipping journals from CA to OA and generally found a slightly higher growth in receiving citations compared to journals in the same discipline and the impact factor's range. However, they didn't observe this trend in all scientific fields.  \cite{momeni2022can} examined the correlation between different factors and the future authors' h-index and found a positive but weak correlation coefficient between them.  

One issue which is often discussed together with OA publishing and APCs is the problem of predatory publishing. Predatory publishers take advantage of the OA movement but work against the good scientific practice. \cite{ross2021dynamics} did a systematic review to study the threat to equity in science via open science implementations. They concluded that less well-resourced researchers, researchers from non-English-speaking countries, and early-career researchers are particularly affected by the ‘predatory publishing’ problem. 

\section{Research questions}
We conduct our study on the association between publishing models, the economic background of researchers, and other author-specific and structural factors along three major research questions:

\textbf{RQ1}: What is the relationship between the income level of researchers’ affiliation countries and their publication behavior (do they prefer OA or CA)?

\textbf{RQ2}: What is the relationship between the income level of researchers’ affiliation countries and their publication behavior (OA or CA) with their citation impact?

To answer these questions, we categorize corresponding authors based on the income level of their affiliation country and compare the access status of articles they have published and their citation impact. Whereas the first two RQs are rather descriptive and aim at quantifying the extent to which access to publish openly and access to read openly (and along with it to make them easier/more likely to cite) are related to the economic background of authors, the third RQ takes a variety of factors into account that have been shown to be strongly associated with tendencies to publish OA \citep{iyandemye2019low,olejniczak2020s,simardgeographic,smith2020assessing,zhu2017support}. 

\textbf{RQ3}: What factors (e.g., journals, articles, authors, or their countries) are associated with selecting the business model of publications (OA against CA)?

Here we aim to give a detailed view of associating factors with OA publishing using correlation, regression and machine learning analyses. To this end, structural features, such as APC waivers, are considered besides author-specific properties, such as gender or years of publishing activity (see Table \ref{tab:features}). We will also look closely at the different access forms to publications such as Gold OA, Hybrid, and Closed Access. Concerning the level of journals, the relationships between journal rankings, APCs, and research fields (Health Sciences, Life Sciences, Physical Sciences, Social Sciences, and multiple fields) will be examined. In addition, possible country-related influencing factors will be investigated, such as countries' income level, transformation agreements' existence, or opportunities for researchers to obtain APC discounts or waivers. At the journal article level, the ratio of OA to CA citations in an article and the number of authors involved are examined. Other author-specific influencing factors can be gender and age, the ratio of OA to CA publications in the past, or even the proportion of international co-authors. 

 \section{Data and methodology}\label{sec1}
To conduct our study, information on the business model, author characteristics, and article impact are needed, and several approaches and databases must be linked to receive a complete dataset. 

\subsection{Data selection}
For the business model of journals (OA, Hybrid, CA)
it is possible to crawl the information from the journal's or publisher's website or to look up sources such as the Directory of Open Access Journals (DOAJ) and Unpaywall, which both include OA information. But information about the history of the business model of journals is rarely available. 
In recent years, many journals have converted (flipped) from closed access to open access and vice versa, but often there is not enough information about the exact date of starting with the new access model. 
The Open Access Directory (OAD), a wiki hosted by the School of Library and Information Science at Simmons University\footnote{http://oad.simmons.edu/oadwiki/Main$\_$Page}, is the only resource containing a list of a few flipped journals and the date of flipping. The open-access start date of journals was available in the DOAJ dataset until 2020. \cite{bautista2020journals} and \cite{momeni2021happens} used OAD and DOAJ for their studies about flipping journals. Unfortunately, DOAJ stopped collecting that information by now: "As time progressed, open access models became more complicated ... It has become harder to find the right answer to that seemingly simple question: when did open access start for this journal?"\footnote{https://blog.doaj.org/2021/02/05/why-did-we-stop-collecting-and-showing-the-open-access-start-date-for-journals/}. \cite{matthias2019two} employed different snapshots of datasets that have the open access status (Scopus, DOAJ, Ulrichsweb, publishers' website, etc.) and some other resources to find out the reverse flip (converting from OA back to CA) and verified them manually.
For the bibliometric analyses related to open access, it is necessary to know about the access status of journals for the period in which we study the effect of OA. Obtaining information more coherently requires looking into different journals' business models and harmonizing them to make them comparable. In addition, every publisher has its own rules for APC exemptions to foster publishing in OA format. For example, eligibility for APC waivers for publishing in Elsevier's journals is based on the 'Research4Life program'\footnote{https://www.research4life.org/access/eligibility/} and for Springer Nature based on 'World bank classification'. Various transformative agreements with publishers and the period of their contracts are other influential factors that should be considered in studying the publishing behavior of each publisher separately.
 
 Due to these varying APC-related rules for different publishers, we focused on one major publisher. To analyse papers for various disciplines and countries, we chose Springer Nature, the largest publisher of academic journals (more than 2,900 journals\footnote{https://www.springernature.com/gp/librarians/products/journals/springer-journals}) with worldwide authors from various disciplines, which provides us with a large amount of data and data diversity for more accurate results. Also, compared to Elsevier, the second most prominent publisher of scholarly journals (above 2,700 journals \footnote{https://www.elsevier.com/about/this-is-elsevier}), this publisher has a higher OA update \citep{sotudeh2015citation,sullo2016open}, resulting in fewer data skewness.
 
\begin{figure}[h]%
\centering
\includegraphics[width=0.5\textwidth]{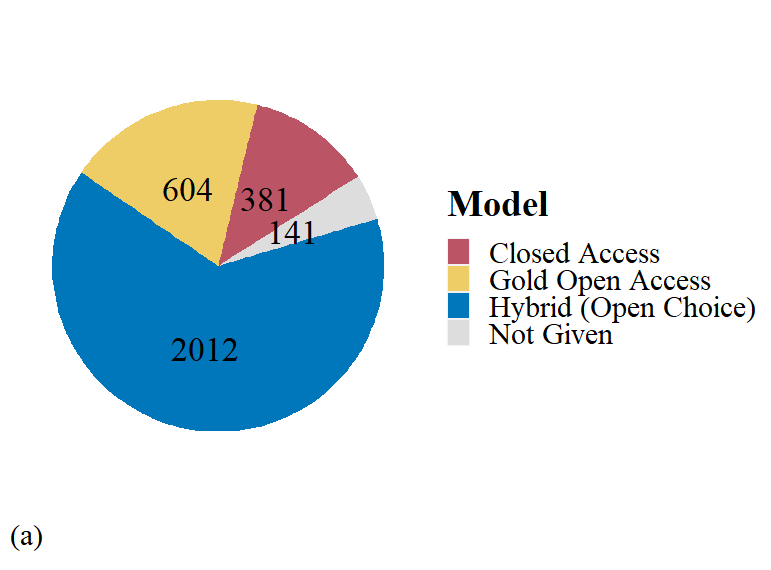}\hfill
\includegraphics[width=0.5\textwidth]{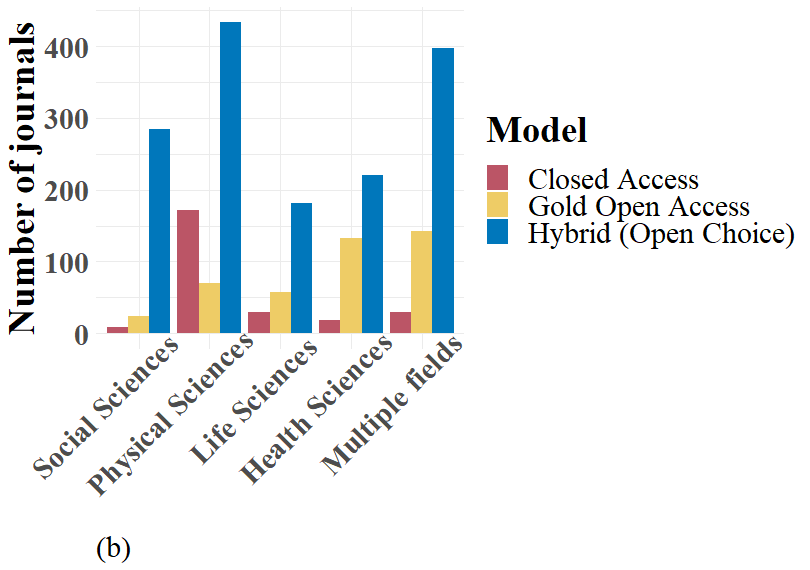}

\caption{Distribution of Springer Nature's journals by (a) publishing model and (b) field and publishing model. }\label{figJrDistr}
\end{figure}

 We downloaded the list of journals and their access status from the snapshot from the year 2019 which is available on the publisher's website\footnote{https://www.springernature.com/gp/open-research/journals-books/journals}. 
 Three publishing models exist for these Springer Nature (SN) journals: Gold Open Access, Hybrid (with the open access option: Open Choice), and Closed Access.  Figure \ref{figJrDistr} displays the distribution of journals and their publishing models. 

For the bibliometric analyses, we employed Scopus\footnote{The in-house Scopus database maintained by the German Competence Centre for Bibliometrics (Scopus-KB), 2021 version}. We matched the list of SN journals with journals in Scopus via title and ISSN. From 3,138 SN journals, we could match 2,757 journals, which we used for further analyses. 
Because of the problems regarding journals' flipping mentioned above, we limited our data to two years, 2017 and 2018, to reduce the errors related to detecting the journals' and articles' business model. It resulted in 522,411 articles.

To detect the publishing model of articles in hybrid journals, we employed Unpaywall\footnote{https://unpaywall.org/} (the snapshot of 2019), a service to find the available version of articles. From metadata in this dataset, we can obtain the publishing model of articles in hybrid journals.

We obtained the APC amount in dollars for 1,741 hybrid journals and 297 gold OA journals from the website of Springer Nature\footnote{{https://www.springernature.com/de/open-research/journals-books/journals}}. There was no fixed APC for 147 gold OA journals (only 5\% of investigated articles belong to these journals), and we had to visit their website to obtain the exact amount for these journals. Therefore we replaced the APC amount for these journals with null values (empty) and excluded them from the data for the classification task. 

To detect the gender status of authors, we utilized a combined name and image-based approach introduced by \cite{karimi2016inferring}, which categorizes the gender into male and female. Based on this method, we tried to detect the gender using the API \textit{Genderize.io} \footnote{https://genderize.io/}. For those names that the API couldn't identify the gender, we looked for names on the web and detected their gender using image-based recognition algorithms which increases the recall and accuracy compared to \textit{Genderize.io} \citep{karimi2016inferring}. We acknowledge that the person's gender is not a binary variable. Considering the social dimensions, more gender identities could not be identified with this approach and that is left out for the analysis. Using Scopus author ID, we found 381,074 unique corresponding authors for the investigated articles, and 10,614 authors (about 3\%) had only initials or no first name, and we couldn't detect their gender.  

Overall, we identified the gender status for 49\% of them. Therefore, we excluded 254,044 articles (about 49\%) that we couldn't detect the gender status of their corresponding author from data in the regression analysis and classification task.
One possible reason for a low rate of identifying gender is the large percentage of authors affiliated with Asian countries (136,591 above 35\%)\footnote{Authors from Armenia, Azerbaijan, Georgia, Kazakhstan, Russia, and Turkey, which belong to both Asia and Europe, are not included in this list.} and probably originally from these countries. Previous studies tested gender detection tools for authors with different nationalities and found them to be less effective for Asian names \citep{karimi2016inferring, santamaria2018comparison}. Table \ref{tab_gender} shows the number and percentage of OA and CA publications belonging to the corresponding authors with a gender status across scientific fields. The percentage of detected gender of authors for OA publications is 4\% more than for CA publications.

\begin{table}[]
    \centering
    \caption{Number and proportion of articles among scientific fields and publishing model that we detected the gender status of their corresponding author.}
    \begin{adjustbox}{width={\textwidth},totalheight={\textheight},keepaspectratio}%
    \begin{tabular}{|c|c|c|}
    \hline
    &\multicolumn{2}{|c|}{\textbf{Publishing Model}}\\
    \hline
   &\textbf{CA model (percentage)}&\textbf{OA model (percentage)}\\
   \hline
Health Sciences &31,642 (53\%) &20,534 (49\%)\\
Life Sciences&23,011 (54\%) &10,032 (57\%)\\
Physical Sciences&74,742 (48\%)& 9,927 (50\%)\\
Social Sciences& 9,210 (40\%)& 2,020 (41\%)\\
Multiple fields&38,507 (52\%)&48,742 (58\%)\\
\hline
Total & 177,112 (50\%) & 91,255 (54\%)\\  
\hline
    \end{tabular}
    \label{tab_gender}
    \end{adjustbox}
\end{table}

\subsection{Features and definitions}
\begin{figure}[!htb]%
\centering
\includegraphics[width=0.9\textwidth]{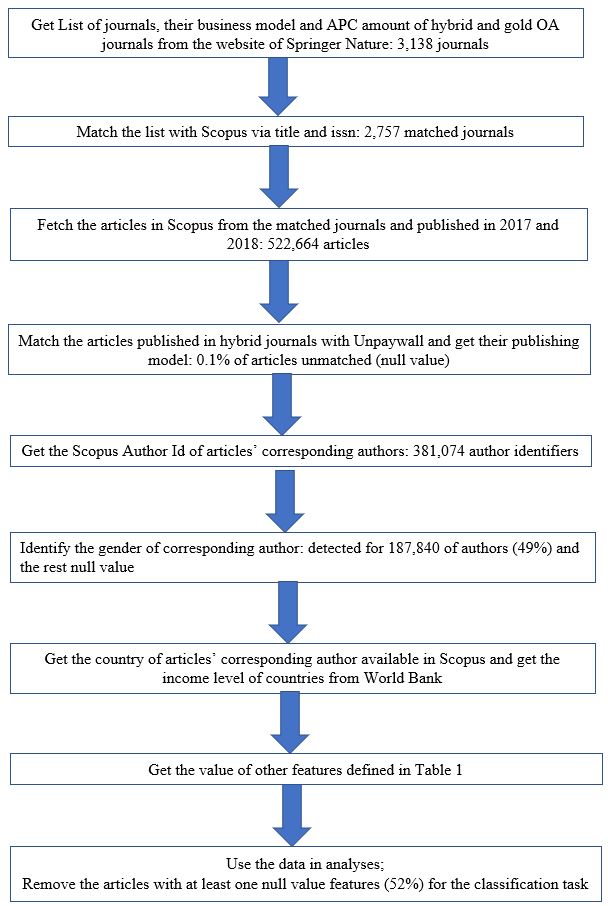}
\caption{Flow chart of data collection and  preparation process.}\label{fig:flowchart}
\end{figure}

To investigate the factors that are associated with higher rates of OA publishing, we defined some features presented in Table \ref{tab:features}. Figure \ref{fig:flowchart} presents an overview of data collection and preparation steps. The final analysed data is available on Git repository \footnote{\url{https://github.com/momenifi/open_access_springer_nature}}.

 To compare the publishing and citation behavior across countries, we classified countries by income based on the World Bank classification\footnote{https://datahelpdesk.worldbank.org/knowledgebase/articles/906519-world-bank-country-and-lending-groups} into four groups: low, lower-middle, upper-middle and high-income economies. The income level of a country has been evaluated every year and its history is available\footnote{http://databank.worldbank.org/data/download/site-content/OGHIST.xlsx}. From 218 listed countries by the World Bank, we excluded 20 countries with different income levels from 2015 to 2018. Springer Nature offers APC waiver and discount to those articles with the corresponding author from low and lower-middle-income countries (classified by the World Bank), respectively\footnote{https://www.springernature.com/gp/open-research/policies/journal-policies/apc-waiver-countries}. 
 
From the website \textit{Transformative Agreement Registry} provided by ESAC\footnote {https://esac-initiative.org/about/transformative-agreements/agreement-registry/} we found three organizations with an open access agreement with this publisher during the investigated years 2017 and 2018 (KEMOE/FWF in Austria, Max Planck Society in Germany and Bibsam consortium in Sweden) and two organizations (VSNU-UKB in Netherlands and FinELib consortium in Finland) in 2018. We obtained the list of involved institutions in the agreement by asking KEMOE/FWF, Bibsam, and  FinELib organizations. The list of participating institutions via VSNU-UK was available on the website of SN \footnote{https://resource-cms.springernature.com/springer-cms/rest/v1/content/19371608/data/v3}. We assumed that the publications with the corresponding author affiliated with institutions included in the transformative agreement are free of APC charges. To find  Max Planck institutions, we used disambiguated institutional addresses for German institutions \citep{rimmert2017disambiguation} available on Scopus-KB. We manually looked up the participating institutions for the rest of the four countries. Altogether, we found 12,323 articles and used them to set the feature 'OA$\_$agreement' value.
 
\begin{figure}[!htb]%
\centering
\includegraphics[width=0.9\textwidth]{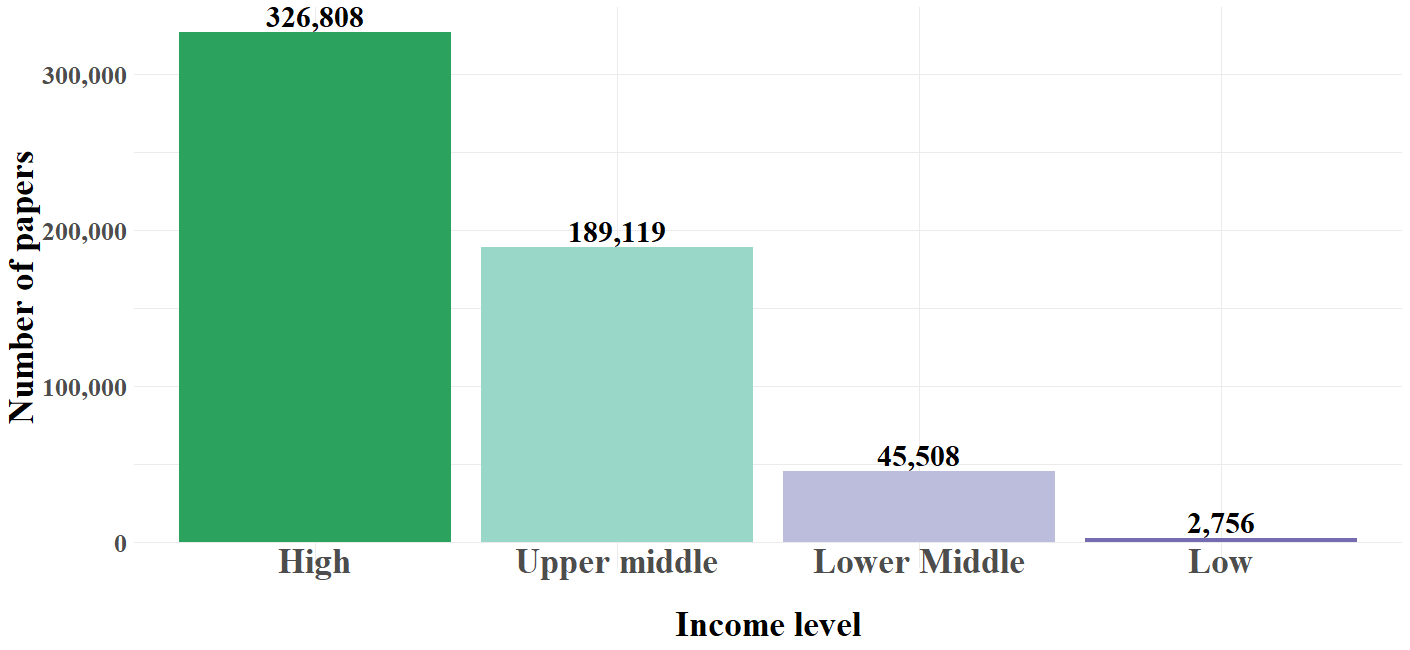}
\caption{Number of papers published by Springer Nature  grouped by income level of countries.}\label{figCntIncmDistr}
\end{figure}

Figure \ref{figCntIncmDistr} represents the number of articles published in Springer Nature in which their corresponding author is affiliated with a country with the respective income group. Sixty-seven articles had a corresponding author with multiple affiliation countries and we excluded them from the analyses. Publication distribution by countries and their income level is available on GitHub\footnote{ https://github.com/momenifi/open$\_$access$\_$springer$\_$nature/blob/main/publications$\_$country$\_$distribution.csv}. 

To obtain the ratio of authors' previous OA publications, we needed to identify authors and their publications. Scopus Author Id enabled us to get each author's list of published articles. For the variable Country$\_$income, we consider average GDP per capita in the years 2017 and 2018 obtained from the world bank group\footnote{https://data.worldbank.org/indicator/NY.GDP.PCAP.CD}. We used the year of the first publication of authors indexed in Scopus to calculate their career age as a measurement of seniority. 

\begin{table}[]
    \centering
    \caption{Features used to study the associated factors with OA publishing.}
    \begin{tabular}{|p{15mm}|l|p{75mm}|}
    \hline
    \textbf{Feature type}&\textbf{Feature}&\textbf{Description} \\
    \hline
\multirow{3}{*}{Journal}&journal\_ranking&H-index ranking of the journal in the related discipline (for multidisciplinary journals, the average ranking among disciplines).\\
&journal\_APC&The cost of APC to publish OA in the journal (US-Dollar).\\
&field&Field of journal (If the journal has more than one field, the value is  \textit{'multiple fields'}).\\
&\hspace{0.3cm}\textit{Health Sciences}&\\
&\hspace{0.3cm}\textit{Life Sciences}&\\
&\hspace{0.3cm}\textit{Physical Sciences}&\\
&\hspace{0.3cm}\textit{Social Sciences}&\\
&\hspace{0.3cm}\textit{multiple fields}& \\
\hline
\multirow{4}{*}{Country}&country\_income&Income level (GDP per capita) of the country in which the corresponding author is affiliated.\\
&OA\_agreement&If the corresponding author's country of affiliation has an OA agreement with the publisher, it equals 1, otherwise 0.\\
&discount\_eligible&If the corresponding author's country of affiliation belongs to the lower-middle income group, it equals 1, otherwise 0.\\
&waiver\_eligible&If the corresponding author's country of affiliation belongs to the low-income group, it equals 1, otherwise 0.\\
\hline
\multirow{2}{*}{Paper}&OA\_cite&ratio of citing OA against CA in this paper\\
&authors\_count&number of authors\\
\hline
\multirow{3}{*}{Author*}&gender&for females equals 1 and for male 0.\\
&age&years since first publication\\
&OA\_publish&ratio of OA publications against CA in the past (number of previous OA publications divided by the number of CA publications)\\
&international\_coauthors&proportion of international co-authors** to all co-authors in this paper\\
\hline
\multicolumn{3}{l}{* Corresponding author}\\
\multicolumn{3}{p{120mm}}{** An international co-author is a co-author who has a different affiliation country than the corresponding author.}\\
    \end{tabular}
    \label{tab:features}
\end{table}

To evaluate the quality of journals and rank them, we employed the journal's H-index, which \cite{hodge2011evaluating}  suggested as a better measurement for ranking journals than the 5-year impact factor in social science and that has been used in previous studies \citep{xia2012positioning,barner2014american}. We calculated the H-index of all journals in Scopus classified in 27 subject categories\footnote{https://service.elsevier.com/app/answers/detail/a\_id/14882/supporthub/scopus/related/1/} within the years 2011 and 2016.

\subsection{Methodology}

\subsubsection{Normalizing the citation impact}
To evaluate and compare the citation impact at the article and journal level among different subject areas, we should normalize them because of varying citation patterns across scientific disciplines and fields. To normalize the journal's H-index across categories, we computed the Percentile Rank (PR) of each journal (inspired by \cite{bornmann2014p100}) in its category. This method gives the journals within a category a rank between 0 (lowest H-index) to 100 (highest H-index). In this approach, journals with the same H-index have the same rank. Therefore, this normalization method is an advantage in case of skewed distributions. If the journal belongs to more than one category, we used the weighted PR \citep{bornmann2020evaluation}. Based on this approach, weighted PR (wPR) will be calculated using the formula:
\begin{equation}
\begin{split}
 wPR = \frac{PR_{sc1} *n_{sc1} +PR_{sc2} *n_{sc2} +...+PR_{sci}*n_{sci}}{n_{sc1} +n_{sc2} +...+n_{sci}}
 \end{split}
 \label{equNorm}
\end{equation}

whereby, $sci$ is the  \textit{i}th subject category that the journal belongs to and $n_{sci}$ is the number of journals in this subject category, and $PR_{sci}$ is PR of the journal in it. 

To present the citation impact of articles, we employed a similar normalizing approach. Because the citation count is confounded by time since publication, we consider the citations during a time window of two years since the publication, as in previous studies
\citep{jannot2013citation,piwowar2018state}. Next, we categorized the articles into groups with the same subject category and publishing year and ranked them from 0 to 100 based on received citations. We define a percentile rank of 50 (citation's median) as a threshold for highly cited articles. An article is highly cited if its rank is above 50\% of PR in its group, meaning that it has received more citations than half of the articles in the same subject category and publishing year.  For articles belonging to multiple subject categories, we used wPR mentioned in Equation \ref{equNorm}, where $sci$ is the  \textit{i}th subject category of the article and $n_{sci}$ is the number of articles in this subject category, and $PR_{sci}$ is PR of the article in it. 
\subsubsection{Correlation analysis}
To find the association between OA publishing and any feature defined in Table \ref{tab:features} we conducted a correlation analysis. The first variable in calculating the correlation is OA publishing, a dichotomous variable (a case of categorical variable). To assess the association with \textit{field}, which is a categorical variable, we selected \textit{Cramer’s V} coefficient. Cramer’s V is based on the chi-squared test and measures the strength of association between two variables. Its value ranges from 0 (no association) to 1 (complete association). The association with binary variables (OA$\_$agreement, discount$\_$eligible, waiver$\_$eligible, gender) was examined with \textit{Phi} coefficient \citep{ekstrom2011phi}. This correlation coefficient ranges from -1 to +1 and shows the strength of the positive or negative correlation between two dichotomous variables. To measure the association with other numerical or continuous variables, we applied the Point-Biserial Correlation Coefficient, which is used instead of the Pearson correlation when a variable is dichotomous \citep{leblanc2017interpretation} and can range from -1 to +1.  

\subsubsection{Regression analysis}
We used multivariate logistic regression to find the relationship between various variables (defined in Table\ref{tab:features}) and OA publishing. It is a common method for modeling the relationship between the dichotomous dependent variable and multiple independent variables. It allows us to understand the association of the dependent variable with an independent variable in the presence of other independent variables in the data. 

\subsubsection{Classification method}
We employed a machine learning method to estimate the likelihood of choosing the publishing model. To this end, we categorized the publishing model of articles into two groups, OA and CA. Then, we utilized the value of defined features in Table \ref{tab:features} to predict the publishing model. This process is a classification task in machine learning.   

To estimate the publishing model of articles, we use a supervised machine learning method, random forest (RF), which is a common tool for classification tasks \citep{kumar2019malware,yamak2016detection,behr2020early,roy2020random}. We utilize this tool for binary classification (OA=1 or CA=0) and use the features introduced in Table \ref{tab:features} as independent variables. We implement the algorithm for hybrid journals in which authors can choose their paper's business model. We used \textit{k}-Fold cross-validation (\textit{k}=10) procedure to train and test the model.

Due to the skewed distribution in the target variable (91\% CA and 9\% OA publishing), we balance them by re-sampling data via \textit{SMOTE} (Synthetic Minority Over-sampling Technique), which was proven to be a suitable method to handle a class imbalance problem \citep{spelmen2018review}. 

\section{Results}\label{sec2}
In this section, first, we present some descriptive statistics about the publishing model of articles across four country groups and address RQ1. Next, we display their differences in terms of citation impact among different models to answer RQ2. Then we focus on RQ3 and present the correlation coefficient between the publishing model and features defined in Table \ref{tab:features} and multivariate logistic regression to show the relationship between variables. Also, we demonstrate the performance of estimating the publishing model of articles in hybrid journals and the importance of defined features in the estimation task to reveal the influential factors in selecting the OA model for publishing. 

\subsection{Countries' income level of corresponding authors and their publishing model}
\begin{figure}[!htb]%
\centering
\includegraphics[width=0.9\textwidth]{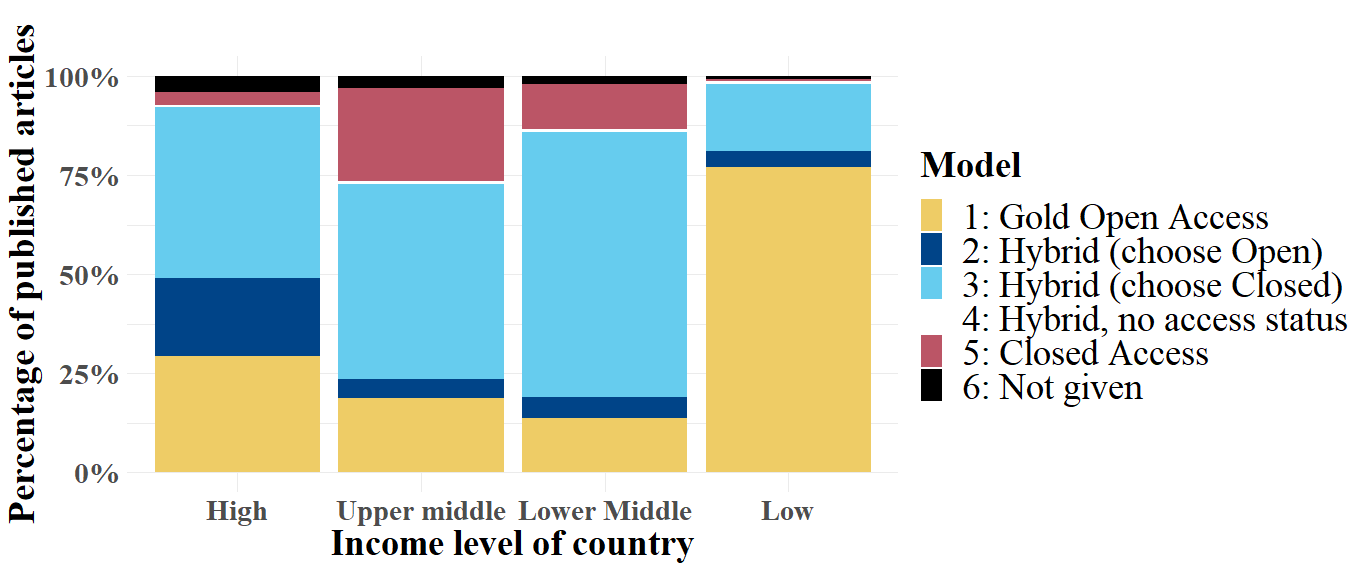}
\caption{Distribution of articles published in journals with three publishing models across four groups of countries. The access status of hybrid articles has been identified from Unpaywall (cases 2 and 3). For case 4 (Hybrid, no access status), we couldn't find hybrid journals' articles in Unpaywall. }\label{fig:publishing_model_articles}
\end{figure}

Figure \ref{fig:publishing_model_articles} shows the distribution of articles categorized by publishing model and the country income level of the corresponding authors. Authors with affiliations in countries with the lowest income level and who are eligible for the APC waiver have the highest proportion of gold OA publications. In contrast to this, authors from lower-middle-income countries who are eligible for the APC discount have the lowest percentage in gold OA publishing.

\subsection{Countries' income level of corresponding authors and their citation impact}
\begin{figure}[!htb]%
\centering
\includegraphics[width=0.9\textwidth]{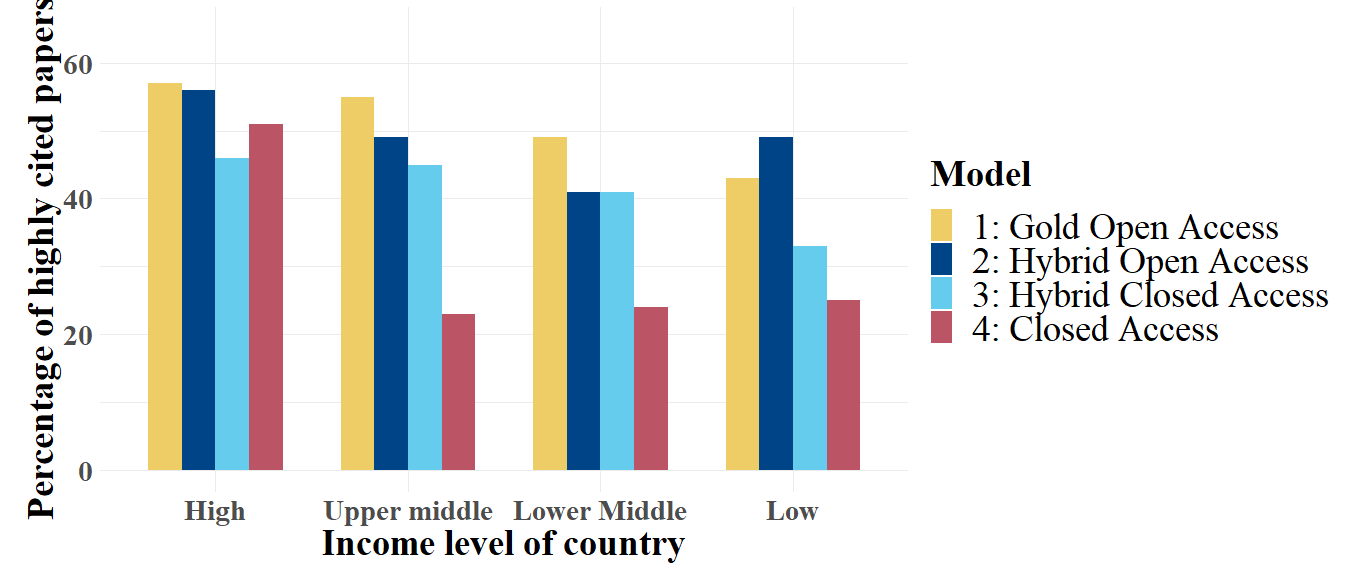}
\caption{Percentage of highly cited papers published in different models. Hybrid Open Access / Closed Access belong to articles published as OA/CA in hybrid journals.}\label{fig_citation}
\end{figure}

 Figure \ref{fig_citation} shows the ratio of highly cited articles for the investigated articles with different publishing models across country groups. Generally, we observe a higher percentage of highly cited papers for corresponding authors from countries with higher income levels. 
 
The ratio of highly cited articles among all countries for gold and hybrid OA models is higher than in other models. Also, this ratio is higher for gold OA articles and indicates the better citation impact of articles published in gold OA journals. The only exception is for countries with low-income levels, with more highly cited papers in the hybrid OA model. Compared to CA journals, journals in hybrid CA have more highly cited articles except for countries with a high-income level. 
 
 \subsection{Influential factors on the publishing model}
 First, we conducted a correlation analysis to find the associations between OA publishing and features. 
 Table \ref{tab_correl} shows the correlation coefficient between the publishing model (if open access is equal to 1 otherwise 0) and features in Table \ref{tab:features}. We separated the data into two sets, set 1 for articles published in OA or CA journals (non-hybrid journals) and set 2 for articles in hybrid journals. Set 1 reveals the association of discount and waiver policies with OA publishing, while optional OA publishing for hybrid journals in set 2 displays more author-specific factors related to OA publishing. The weak negative correlation with \textit{gender} demonstrates that the tendency toward gold OA publishing for women is slightly more than for men, which disagrees with previous findings \citep{zhu2017support,olejniczak2020s}. As we observed the lowest proportion of OA publishing for countries with a lower-middle-income level in Figure \ref{fig:publishing_model_articles}, the negative correlation for \textit{discount$\_$eligible} (also positive value for \textit{waiver$\_$eligible}) in Table \ref{tab_correl} points out that the discount policies are insufficient to motivate the authors from these countries for gold OA publishing. 
 Table \ref{tab_regression} displays the relationship between the publishing model and features in Table \ref{tab_correl} by considering all features in multivariate logistic regression. The results confirm the negative/ positive correlation calculated in correlation analysis, except the positive correlation between \textit{discount$\_$eligible} and the publishing model is inconsistent with the result in the correlation coefficient. The highest Odds Ratios for Social Sciences among fields in Table \ref{tab_regression} reveal the highest proportion of OA publishing in this field. This field has experienced a dramatic growth of OA journals since 2009 \cite{liu2018open}.
 The strong positive correlation between \textit{journal$\_$ranking} and the publishing model for the first set suggests that the journal's rank is the dominant factor in choosing a gold OA journal to publish. Therefore, we estimate the publishing model for articles in set 2 (hybrid journals) to discover other feature categories rather than journal-speciﬁc factors inﬂuencing the authors' decision for an OA option. Moreover, the optional choice of the OA model in hybrid journals better reveals characteristics leading to the OA model.

\begin{table}[!htb]%
    \centering
        \caption{Correlation coefficient between independent variables and the target variable. The value of the target equal to 1 (0) means the paper has been published in the OA (CA) model. }
    \begin{tabular}{|l|l|c|c|}
    \hline
    \multicolumn{2}{|c|}{}&\multicolumn{2}{c|}{\textbf{Correlation Coefficient}}\\
    \hline
    \textbf{Feature}&\textbf{Correlation Test}& \textbf{Set 1 (non-hybrid)}& \textbf{Set 2 (hybrid)}\\
    \hline
\hspace{0.1cm}journal\_ranking&Point-Biserial&0.70&0.07\\
\hspace{0.1cm}journal\_APC&Point-Biserial&-&0.10\\
\hspace{0.1cm}field&Cramer’s V&0.69&0.09\\
\hline
\hspace{0.1cm}country\_income&Point-Biserial&0.28&0.16\\
\hspace{0.1cm}OA\_agreement&Phi&0.08&0.30\\
\hspace{0.1cm}discount\_eligible&Phi&-0.08&-\\
\hspace{0.1cm}waiver\_eligible&Phi&0.06&-\\
\hline
\hspace{0.1cm}OA\_cite&Point-Biserial&0.42&0.13\\
\hspace{0.1cm}authors\_count&Point-Biserial&0.09&0.07\\
\hline
\hspace{0.1cm}gender&Phi&-0.08&-0.01\\
\hspace{0.1cm}age&Point-Biserial&-0.08&0.02\\
\hspace{0.1cm}OA\_publish&Point-Biserial&0.46&0.41\\
\hspace{0.1cm}international\_coauthors&Point-Biserial&0.17&0.11\\
\hline
\multicolumn{2}{|c|}{\textbf{Sample Size:}}&192,498 &329,913\\
\hline
    \end{tabular}
    \label{tab_correl}
\end{table}

\begin{table}[!htb]%
    \centering
        \caption{The results of Logistic regression. The target variable is the publishing model and is equal to 1 for OA and 0 for CA publishing. The outputs are Odds Ratio, $\exp(\beta)$. (1-$\exp(\beta)$) shows the percentage change of the target variable per unit increase in an independent variable. So, the Odds Ratio greater/less than one displays a positive/negative correlation between variables.}
 \begin{adjustbox}{width={\textwidth},totalheight={\textheight},keepaspectratio}%
    \begin{tabular}{|l|c|c|c|c|}
    \hline
    &\multicolumn{2}{|c|}{Set 1}&\multicolumn{2}{|c|}{Set 2}\\
    \hline
    &Odds Ratio&95$\%$ CI&Odds Ratio&95$\%$ CI\\
    \hline
\textbf{Intercept}&0.002\threeS(-72.4)&0.001 to 0.002&0.00\threeS(-87.7)&0.00 to 0.00\\
\textbf{Independent Variables}&&&&\\
\hspace{0.1cm}journal\_ranking&1.98\threeS(10.38)&1.74 to 2.25&110.7\threeS(86.5)&99.5 to 100.23\\
\hspace{0.1cm}journal\_APC&1.00\threeS(8.05)&1.0001 to 1.0002&-&-\\
\hspace{0.1cm}field&&&&\\
\hspace{0.3cm}\textit{Health Sciences}& reference& reference&reference&reference\\
\hspace{0.3cm}\textit{Life Sciences}&1.01(0.31)&0.94 to 1.08&0.67\threeS(-9.55)&0.62 to 0.73\\
\hspace{0.3cm}\textit{Physical Sciences}&0.97(-0.91)&0.91 to 1.07&0.20\threeS(-44.29)&0.18 to 0.21\\
\hspace{0.3cm}\textit{Social Sciences}&1.90\threeS(13.81)&1.73 to 2.08&3.49\threeS(12.2)& 2.86 to 4.27\\
\hspace{0.3cm}\textit{multiple fields}&1.25\threeS(8.5)& 1.19 to 1.32&3.4\threeS(30.87)& 3.17 to 3.71\\
\hspace{0.1cm}country\_income&1.00\threeS(33.88)&1.000 to 1.000&1.000\threeS(16.18)&1.00 to 1.00\\
\hspace{0.1cm}OA\_agreement&14.9\threeS(65.07)&13.78 to 16.22&0.93(-0.78)&0.78 to 1.11\\
\hspace{0.1cm}discount\_eligible&-&-&1.7\threeS(9.17)&1.52 to 1.90\\
\hspace{0.1cm}waiver\_eligible&-&-&20.19\threeS(5.53)&8.29 to 77.5\\
\hspace{0.1cm}OA\_cite&0.55\threeS(-12.97)&0.500 to 0.600&1.55\threeS(8.4)&1.39 to 1.71\\
\hspace{0.1cm}authors\_count&1.003(0.80)&0.99 to 1.01&1.17\threeS(33.15)&1.16 to 1.18\\
\hspace{0.1cm}gender&0.94\twoS(-2.8)&0.90 to 0.98&0.93\oneS(-2.5)&0.88 to 0.98\\
\hspace{0.1cm}age&1.05\threeS(29.63)&1.05 to 1.1.054&0.97\threeS(-15.36)&0.96 to 0.98\\
\hspace{0.1cm}OA\_publish&196.79\threeS(105.65)&178.46 to 217.09&23.86\threeS(50.58)&21.1 to 26.99\\
\hspace{0.1cm}international\_coauthors&1.17\threeS(18.21)&1.15 to 1.19&1.03(1.34)&0.99 to 1.06\\
\hline
 \textbf{McFadden's Pseudo} \boldmath{$R^2$}&\multicolumn{2}{|c|}{0.25}&\multicolumn{2}{|c|}{0.60}\\
\hline
  \textbf{Sample Size} &\multicolumn{2}{|c|}{96,674}&\multicolumn{2}{|c|}{162,773}\\
 
\hline
\multicolumn{5}{|l|}{significant codes: . $p<0.1$, $\oneS p<0.05$, $\twoS p<0.01$, $\threeS p<0.001$}\\
\multicolumn{5}{|l|}{z-values of coefficients in parentheses}\\
\multicolumn{5}{|l|}{CI: Confidence Interval}\\

\hline
    \end{tabular}
    \label{tab_regression}
    \end{adjustbox}
\end{table}

\begin{table}[!htb]
    \centering
        \caption{performance of predicting the publishing model of papers with random forest method.}
    \label{tab:performance}
    \begin{tabular}{|c|c|c|}
    \hline
    \textbf{Classification}  &\textbf{OA}&\textbf{CA}\\
    \hline
    Precision&0.85 &0.94\\
    Recall&0.95&0.83\\
    F1score&0.89&0.88\\
    \hline
    Accuracy&\multicolumn{2}{|c|}{0.92}\\
         \hline
    \end{tabular}
\end{table}

\begin{figure}[!htb]
    \centering
    \includegraphics[width=0.9\linewidth]{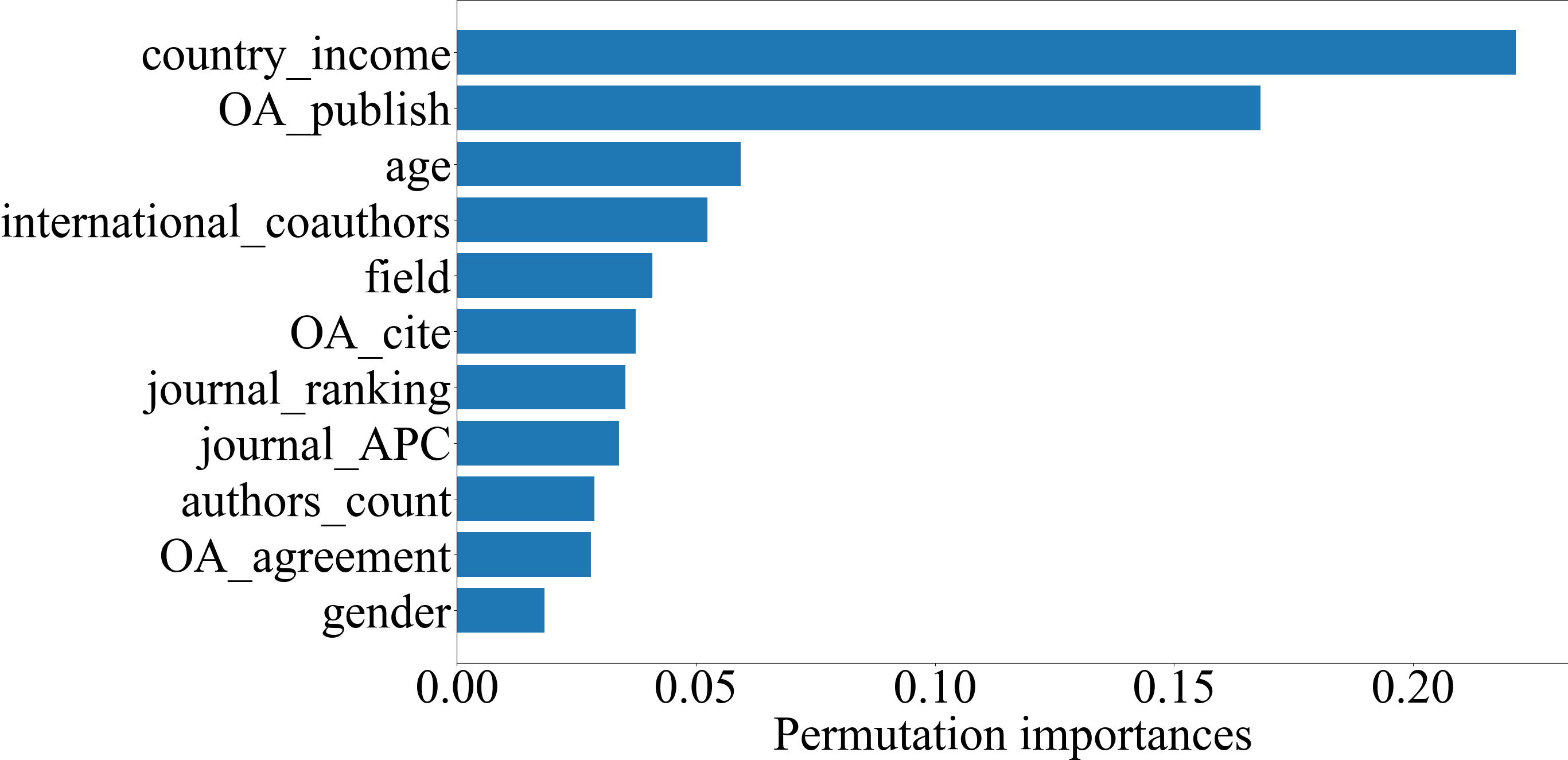}\hfill
    \caption{Permutation importance of features employed to predict the publishing model of papers with random forest method for the articles published in hybrid journals. }
    \label{fig:importance}
\end{figure}

Table \ref{tab:performance} shows the performance of the  RF classifier for the second set (hybrid journals). Figure \ref{fig:importance} displays the \textit{permutation importance} of features employed to predict the publishing model implemented for this set. The permutation importance of a feature shows a decrease in the model performance when the feature's value is randomly shuffled while the values of other predictors remain unchanged. A higher value for a feature shows more predictive power in the proposed model. The highest importance values for \textit{country$\_$income}, and \textit{age} in Figure \ref{fig:importance} indicate that the most significant factors in selecting an OA model are the income level of countries and seniority. The lowest value for the variable \textit{gender} presents that gender has a lower impact on the authors' decision for the OA model compared to other factors. OA$\_$agreement is one of the weakest features in predicting the publishing model, and the correlation analysis also shows a weak correlation between them. 
One possible reason for the weak effect is that only 2.3\% of papers have been involved in transformative agreements. In addition, the income level of countries is the most important feature, and regarding the positive correlation of this feature with OA publishing, it is more likely for authors from high-income countries (even without a transformative agreement) to publish in the OA model. This may also smooth the association of the agreement with OA publishing.

\section{Conclusion and discussion}\label{sec13}
This work presents a detailed study of the relationship between author-specific and structural factors (e.g., income level of authors' affiliation country), OA publishing, and OA citation advantage. First, we investigated the relationship between the income level of countries and OA publishing for articles published by Springer Nature in the years 2017 and 2018. We found that authors from lower-middle-income countries with the eligibility to use APC discounts have a lower proportion of gold OA publications in all published papers by this publisher compared to other countries. It indicates that discounted APC is still too much for these authors to pay for a gold OA model and agrees with the statement of \citep{rouhi2022left}, who pointed out that waiver and discount issues couldn't bring author equity in reading and publishing. In contrast, this proportion of authors from countries with a low-income level who receive APC waivers is higher than authors from other countries. This result conflicts with the study's results by \cite{smith2020assessing}, which found fewer OA papers proportions published by Elsevier for these countries compared to others. The reason can be stricter conditions, which this publisher considers for waiver eligibility.

We examined the citation impact of these articles and compared the percentage of highly cited papers among the publishing models and the income levels of the corresponding authors' countries. For all countries, the OA model in gold OA or hybrid has the highest percentage of highly cited papers. Also, the results demonstrate a higher proportion of highly cited articles for countries with higher income levels. Although it displays more citation impact for OA models, it can result from confounding factors such as self-selection and  quality biases \citep{gargouri2010self}. Also, examining the preprint  and green OA publishing (if the article has been published in the CA model, but a free version is available in a repository outside of the publisher's website) effect will result in more accurate analyses \citep{fraser2020relationship,wang2020impact}.

To find more characteristics (e.g., author, journal, paper) related to OA publishing, we conducted correlation, regression, and machine learning analyses. The results of the correlation analysis displayed the strength of positive/negative correlation between the publishing model and every feature defined in Table \ref{tab:features}. Using regression analysis, we examined the association of each factor while accounting for other factors. The results reinforced the correlation outcomes. The only conflict between these two methods was the negative correlation between \textit{discount$\_$egibility} with  OA publishing in correlation analysis, but positive in regression evaluation. In addition, we estimated the publishing model of articles (OA or CA) using a random forest-based machine learning approach and examined the impact of each feature on the estimation task. 
The results show that the country's income and more experiences in OA rather than CA publishing are the most influential factors in estimating the publishing model. We discovered that the tendency toward OA publishing was slightly higher for women, but it was a less important feature than other features in estimating the OA model.

\section{Limitations and future work}\label{sec12}
One obvious limitation of this study is that we included articles from just one publisher, Springer Nature. Authors' publishing behavior may differ among articles published by other publishers, which limits the generalizability of the results of our study.  

We obtained the access status of journals in 2019 based on the list published on Springer Nature's website (the same for the access status at the article level from Unpaywall). Some journals may have flipped from CA to OA \citep{momeni2021happens} or vice versa, and we did not detect them, which can cause errors in results. Furthermore, we did not control the correctness of external data (Springer nature and Unpaywall). The accuracy of these data affects the results' precision. We identified the gender of 49\% authors and 
 removed 49\% of articles without gender status corresponding authors in regression and machine learning analyses. In addition, 2\% of the data have been removed because of the null value in other features (e.g., journals' APC). Because the gender detection approach doesn't work well for Asian names, especially Chinese ones, we have a lower proportion of these authors with gender status in the dataset, which also creates biases in our analyses. 

 For future work, we can consider other publishers to examine how the different APC policies among publishers impact OA publishing. Also, controlling for articles' language in the analyses encourages future studies. \textbf{Springer Nature} is an international publisher and publishes mostly articles in English\footnote{https://support.springernature.com/en/support/solutions/articles/6000219817-are-any-of-your-titles-available-in-other-languages-}, and articles in other languages are underrepresented in this study. considering other publishers with non-English content and the articles' language in the analyses can reveal the role of languages in publishing international OA articles and citation advantages.

\section{Declarations}

\subsection*{Author contributions}
\textbf{Fakhri Momeni:}  Conceptualization; Methodology; Software; Validation; Formal analysis; Investigation; Resources; Writing - Original Draft; Writing - Review $\&$ Editing; Visualization.\\
\textbf{Kristin Biesenbender:} Conceptualization; Resources; Writing - Review $\&$ Editing. \\
\textbf{Philipp Mayr:} Writing - Review $\&$ Editing; Project administration; Funding acquisition.  \\
\textbf{Stefan Dietze:} Supervision; Methodology; Writing - Review $\&$ Editing \\
\textbf{Isabella Peters:} Supervision; Writing - Review $\&$ Editing; Project administration; Funding acquisition;

\subsection*{Competing interests}
  The authors declare that they have no competing interests.

\subsection*{Availability of data and materials}
The dataset analysed during the current study and codes are available on the \href{https://github.com/momenifi/open_access_springer_nature.git}{Git repository}.

\subsection*{Acknowledgements}
This work is financially supported by BMBF project OASE, grant number 01PU17005A. We acknowledge
the support of the German Competence Center for Bibliometrics (grant: 01PQ17001) for maintaining the used dataset for the analyses.

\bibliography{sn-bibliography}


\end{document}